\documentclass[aps,pra,superscriptaddress,notitlepage,floatfix]{revtex4}
\usepackage{amssymb, graphicx,subfigure}
\usepackage{enumerate}
\usepackage{amsmath,bm}
\usepackage{dcolumn}
\usepackage{color}
\usepackage{natbib} 
\usepackage[latin1]{inputenc}

\begin{document}
\newcommand{\ket}[1]{|#1\rangle}
\newcommand{\bra}[1]{\langle #1|}
\newcommand{\escalar}[2]{\langle #1|#2\rangle}
\newcommand{\expectation}[3]{\langle #1|#2|#3\rangle}
\newcommand{\expected}[1]{\left\langle #1\right\rangle}

\title{Rotational spectrum of asymmetric top molecules in combined static and
laser fields}

\author{J.\ J.\ Omiste}

\author{R.\ Gonz\'alez-F\'erez}

\affiliation{Instituto Carlos I de F\'{\i}sica Te\'orica y Computacional,
and Departamento de F\'{\i}sica At\'omica, Molecular y Nuclear,
  Universidad de Granada, 18071 Granada, Spain} 

\author{P.\ Schmelcher}

\affiliation{Zentrum f\"ur Optische Quantentechnologien, Luruper Chaussee 149, Universit\"at
  Hamburg, 22761 Hamburg, Germany} 

\date{\today}
\begin{abstract}
We examine the impact of the combination of a static electric field and a non
resonant linearly polarized laser field on an asymmetric top molecule. 
Within the rigid rotor approximation, we 
analyze the symmetries of the Hamiltonian for all possible field configurations.
For each irreducible representation, the Schr\"odinger equation is solved by a
basis set expansion in terms of a linear combination of Wigner
  functions respecting the corresponding symmetries, which 
allows us to distinguish avoided crossings from genuine ones.
Using the fluorobenzene and pyridazine molecules as prototypes,
the rotational spectra and properties are analyzed for experimentally accessible
static field strengths and laser intensities.
Results for energy shifts, orientation, alignment and hybridization of the angular 
motion are presented as the field 
parameters are varied.  
We demonstrate that a proper selection of the fields gives rise to a constrained
rotational motion in the three Euler angles, the wave function being oriented along 
the electrostatic field direction, and 
aligned in the other two angles. 
\end{abstract}

\maketitle

\section{Introduction}

The manipulation of large %polyatomic 
molecules by using external fields represents, in spite of its long history, a
very active and promising research area.
 %during the last years, 
Indeed, major efforts have been undertaken to
create samples  of oriented and/or aligned molecules, and
a large variety of experimental techniques have been developed, 
such as, e.g., the brute force orientation~\cite{loesch:jcp93}, 
hexapole 
focusing~\cite{brooks:science,parker:1989,hain:jcp111}, a train of laser
pulses~\cite{lee:phys_rev_lett_97, poulsen:phys_rev_a_73},
or  
a combination of a laser pulse and a weak static electric 
field~\cite{kupper:prl102,kupper:jcp131,nevo:pccp11}. 
The control and manipulation of the directional features  of molecules, i.e.,
of their 
rotational degree of freedom, optimize the information content on 
experimental measurements performed in  the laboratory frame.
Indeed, the availability of asymmetric top  molecules in oriented and/or
aligned pendular states allows for a wealth of interesting applications in
areas as diverse as spectroscopy~\cite{moore:jcp110,kanya:j_chem_phys_121},
photoelectron angular 
distributions~\cite{Holmegaard:natphys6,PhysRevA.83.023406},
sterodynamic control of chemical 
reactions~\cite{aquilanti:pccp_7,brooks:jcp45,aoiz:chem_phys_lett_289, zare:science}, 
dissociation of 
molecules~\cite{wu:jcp101,baumfalk:jcp114,brom:11645,lipciuc:123103}, 
electron diffraction~\cite{PhysRevLett.102.213001},
or high-harmonic 
generation~\cite{PhysRevLett.99.243001,velotta:phys_rev_lett_87}.

The experimental achievements have been accompanied by 
theoretical efforts to understand and explain the intriguing physical
phenomena appearing in asymmetric top molecules exposed to external fields. 
Regarding the impact of radiative fields on these molecules, 
the corresponding theoretical studies have been especially fruitful in
explaining a vast amount of experimental results, such as, e.g.,   
the rotational revival structure following the irradiation by an intense
picosecond  
laser pulses~\cite{poulsen:jcp121}, 
the three-dimensional alignment by elliptically polarized laser 
fields~\cite{larsen:phys_rev_lett_85_2470,artamonov:j_chem_phys_128,artamonov:phys_rev_a_82_023413}
or the use of  long and short laser pulses to control the 
rotation~\cite{viftrup:prl99,viftrup:pra79}.
Analogously, the motivation of the theoretical works considering an
electrostatic  field was either 
to interpret some 
experimental results~\cite{moore:jcp110}, or to confirm the feasibility of
other experiments, e.g., the Stark deceleration of 
polyatomic asymmetric molecules~\cite{schwettman:jcp123}. 
The molecular orientation due to the interaction with a static electric field
has been investigated for asymmetric top molecules with their permanent dipole
moment  $ \boldsymbol{\mu}$ 
parallel to a principal axis of inertia, and for the non-parallel 
case~\cite{bulthuis:jpca101,hain:jcp111,Kong:jpca104}.
In the strong electrostatic field regime, an analytical study
of the energy-level representation has shown that the asymmetric top pendular  
states are well described by a two-dimensional anisotropic harmonic
oscillator~\cite{kanya:pra70}, and it has been used to
reproduce spectroscopic results in the pendular regime for static fields up to
$200$ kVcm$^{-1}$~\cite{kanya:j_chem_phys_121}.  

A detailed analysis of the rotational spectrum of symmetric top molecules 
exposed to combined electrostatic and nonresonant radiative fields was
recently  performed by  
H\"artelt and Friedrich~\cite{hartelt_jcp128}. For tilted fields, 
only the projection of the total angular momentum  $\mathbf{J}$ onto the
body fixed frame $z$-axis $K$ remains as good quantum number, and 
a 2D description of the rotational spectrum of the molecule is required.
The corresponding dynamics is very complicated, indeed, 
in the presence of a static electric field, 
it has been shown that the molecular spectrum presents classical and 
quantum monodromy~\cite{kozin:jcp118}. 
They provide correlation diagrams between the field-free states and the
pendular  levels of the intense laser field~\cite{kim:jcp108}
as well as the strong electrostatic field regime.
For a selection of states, they investigate the energy shifts and directional 
properties (orientation and alignment) for parallel and perpendicular fields.
In these systems, the coupling of both field interactions could  provoke an
enhancement of the orientation giving rise to an 
oriented and antioriented pair of levels.    
For an oblate system, this phenomenon appears 
in the tunneling doublets created by the interaction of the molecular 
polarizability with the linear polarized laser field
(this effect  was already analyzed for linear 
molecules~\cite{friedrich:jpca103,friedrich:jcp111,cai:prl01}), whereas for a
prolate 
molecule, it appears among exactly degenerate doublets of indefinite parity
appearing in the strong laser field regime.

A classical theoretical analysis of asymmetric top  molecules 
exposed to a combination of static and laser fields has been 
performed recently~\cite{arango:ijbc1127}. 
However, the quantum analog has not
yet been addressed in the literature, to the best of our knowledge. 
Recently, the authors have developed a diabatic model to describe the
evolution of 
alignment and orientation of asymmetric top molecules in combined 
fields as the laser intensity is varied~\cite{omiste:2011}. 
The outcome of this theoretical study has been 
compared to the experimental data obtained for
the benzonitrile molecule~\cite{PhysRevA.83.023406} proving 
the importance of non-adiabatic processes in the field-dressed molecular
dynamics.
Thus, motivated by the current experimental interest on these asymmetric molecules~\cite{kupper:prl102,kupper:jcp131,nevo:pccp11,Holmegaard:natphys6,PhysRevA.83.023406}
and by the fact that the rotational dynamics of 
most polyatomic molecules can be described as asymmetric tops, 
we extend in the present work the previous study on symmetric 
tops~\cite{hartelt_jcp128} to these more complicated 
systems.
We perform a theoretical investigation of  an asymmetric top in the
presence of combined electrostatic and nonresonant radiative fields
within the rigid rotor description. 
The field-dressed rotational spectrum is significantly more complicated,  
and the more general case of no collinear field requires a full 3D
description. 
We will perform a detailed analysis of the symmetries  of the Hamiltonian for
all  possible field configurations.
In tilted fields, the reduction of the
symmetries enhances the complexity of the spectrum, and 
a large amount of avoided crossing appears
between states of the same symmetry.
Hence, to simplify the analysis and interpretation of our 
results, 
the Schr\"odinger equation  is numerically  solved 
for each irreducible representation 
by expanding the wave function in a basis with the corresponding symmetry.
As prototype examples we consider (C$_6$H$_5$F) and pyridazine
(C$_4$H$_4$N$_2$) molecules.  
These two systems have similar values of their polarizability
tensors and dipole moments, but different inertia tensors and are, therefore,
affected differently by the external fields.  
We explore their rotational spectrum as either
the laser intensity, the electrostatic field strength, or the  inclination angle
between them is varied. Our focus is on the energy shifts, the directional 
properties and the hybridization of the angular motion. 
Depending on the dominant interaction, a rich field-dressed dynamics 
is observed with levels achieving different degrees of orientation  and/or
alignment. The role played by the inclination angle is exemplary investigated
via a set of states and in avoided crossings between two adjacent
levels. Moreover, we show that due to the combination of  both field
interactions the rotational motion is restricted in the three Euler angles,
being oriented along the static electric field direction and constrained
in the $XY$ plane of the laboratory frame, which is perpendicular to the laser
polarization. 
This mechanism of orientation and 2D alignment is very sensitive to 
the field parameters and to the molecular properties.
 
The paper is organized as follows. In Sec. \ref{sec:hamiltonian_system}  the
rotational Hamiltonian  is presented together with  a comprehensive 
consideration of its symmetries for the different field configurations. In
Sec. \ref{sec:results}, we discuss the numerical  results for two asymmetric
molecules, fluorobenzene and pyridazine, as the field parameters are modified.
In particular,  we explore three different cases: 
i) for fixed laser intensity and three 
inclination angles, we vary
the electrostatic field strength; 
ii) for fixed electrostatic field and three 
inclination angles the laser intensity is enhanced; 
and iii) for fixed laser intensity and electrostatic field strengths,  
the angle between 
them is continuously changed from $0$ to $\pi/2$.  The conclusions and outlook are
provided in Sec. \ref{sec:conclusions}. 

\section{Hamiltonian of an asymmetric top molecule in the presence of the fields}
\label{sec:hamiltonian_system}

\begin{figure}[b]
\begin{center}
 \includegraphics[scale=.2]{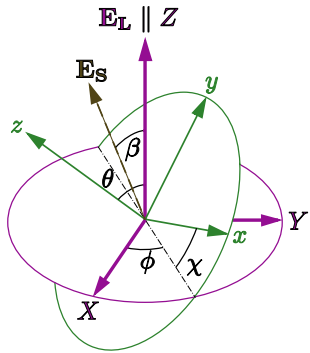}
\end{center}
\caption{Laboratory and molecular fixed coordinate frames and
  field configuration.} 
\label{fig:fields}
\end{figure}

We consider a polar and polarizable asymmetric top molecule exposed to an
homogeneous static electric field and a nonresonant  
linearly polarized laser. Our study is  restricted to the regime of
field strengths that significantly affects the rotational dynamics of the
molecule, whereas its impact on the electronic and vibrational  
structure can be described by first order perturbation theory.
We work within the Born-Oppenheimer approximation, assuming that the
rotational and vibrational dynamics can be adiabatically separated, and 
apply a rigid rotor description of the molecular systems.
Furthermore, we neglect relativistic, fine and hyperfine
interactions as well as couplings of different electronic states.  
In the laboratory fixed frame (LFF) $(X,Y,Z)$, the $Z$-axis is chosen parallel
to the polarization of the laser, and the direction of the  
homogenous electric field is taken forming an angle $\beta$ with this axis and
contained in the $XZ$ plane. 
The molecular or body fixed frame (MFF) $(x,y,z)$
is defined so that the permanent electric dipole moment is parallel to the
$z$-axis, and for the considered systems the smallest moment of inertia is 
parallel to the $x$-axis. 
The relation between both frames is given by the Euler angles 
$\Omega=(\phi,\;\theta,\;\chi)$~\cite{zare}, which are shown together with  
the field configurations in Fig.~\ref{fig:fields}. 
We only analyze molecules having the electric dipole moment parallel to one
of the axis, and a diagonal polarizability tensor. 
Thus, the rigid rotor  Hamiltonian reads
\begin{equation}
 H=H_r+H_S+H_L,
\label{eqn:hamiltonian}
\end{equation}
where $H_r$ is the field-free Hamiltonian, and $H_S$ and $H_L$ stand for
the interactions with the static and the laser field, respectively.  

In the absence of the fields, the  rigid rotor Hamiltonian is given by
\begin{equation}
 H_r=B_xJ_x^2+B_yJ_y^2+B_zJ_z^2
\label{eqn:hamiltonian_rot}
\end{equation}
where the angular momentum operators refer to the MFF, with $B_i=\hbar^2/2I_{ii}$ being the rotational constant
and $I_{ii}$ the moment of inertia around the principal axis of inertia $i$,
with $i=x,y$ and $z$. An estimation of the degree of asymmetry is provided by Ray's parameter,
$\kappa=(2B_y-B_z-B_x)/(B_y-B_x)$, where %$B_z$, $B_y$ and $B_x$  
the rotational constants of the considered system satisfy $B_z\ge B_y\ge B_x$. 
For a symmetric top $(B_x=B_y)$, Ray's parameter
takes the extreme values 
$\kappa=1$ and $-1$ for the oblate and prolate cases, respectively.

For the regime of electrostatic field strengths $\mathbf{E_S}$ considered here, we can neglect 
the interaction 
via the molecular polarizability, arriving at the following the Stark Hamiltonian
\begin{equation}
 H_S=-\mathbf{E_S}\cdot \boldsymbol{\mu}=-E_S\mu \cos\theta_S,
\label{eqn:hamiltonian_static}
\end{equation}
where  $\theta_S$ is the angle between the permanent molecular 
electric dipole moment, 
$\boldsymbol{\mu}=\mu\mathbf{\hat z}$, and the static electric field with 
$\cos\theta_S=\cos\beta\cos\theta+\sin\beta\sin\theta\cos\phi$ and 
 $0\le\beta\le\pi/2$. 

Here, we consider  a nonresonant 
laser field linearly polarized along the $Z$ axis, $\mathbf{E_L}(t)=E_{\max}g(t)\cos(2\pi \nu t) \mathbf{Z}$, with 
the frequency $\nu$, the field strength $E_{\max}$, and $g(t)$ being the pulse 
envelope. We assume that $\nu^{-1}$ is much shorter than the pulse duration or
the rotational period, so that we can average over the rapid oscillations, which  causes the coupling of this field with the permanent dipole moment to
vanish~\cite{dion_pra59,henriksen_cpl312}. In addition, 
we assume that the pulse duration is much longer than the rotational period of
the molecular system, such that the states adiabatically follow the change of the
field, and we restrict our analysis to the adiabatic limit $g(t)\to 1$.  
Hence, the interaction of the laser with the polarizability is the leading
order term, and can be written as 
\begin{equation}
 H_L=-\cfrac{I}{2c\epsilon_0}
(\alpha^{zx}\cos^2\theta+\alpha^{yx}\sin^2\theta\sin^2\chi) 
\label{eqn:hamiltonian_laser}
\end{equation}
with $\alpha^{ji}=\alpha_{jj}-\alpha_{ii}$, and $\alpha_{ii}$ being the $i$-th
diagonal element of the polarizability tensor, 
with $i=x,\;y,\;z$~\cite{seideman}. In this expression we have used $\langle E_L^2\rangle= I/c\epsilon_0$,
$c$ being the speed of light,
$\epsilon_0$ the dielectric constant,
and expectation value $\langle E_L^2\rangle$ indicates the time average. 

Our aim is to investigate the rotational spectrum of an asymmetric top
molecule exposed to different field configurations. To do so, we solve the
time-independent Schr\"odinger equation associated to the Hamiltonian
\eqref{eqn:hamiltonian}, which even in the field-free case can not be
 solved analytically. Let us start by analyzing how the symmetries of this
Hamiltonian change when the angle between the fields varies.

\subsection{Symmetries}
\label{sec:symmetries}
A detailed analysis of the symmetries of an asymmetric top rotor in the
field-free case and exposed to a static
electric field has been performed in Ref.~\onlinecite{kanya:pra70}. Here, we extend
this study to the  
field configurations investigated in this work.

The symmetries of the field-free Hamiltonian \eqref{eqn:hamiltonian_rot}
are the spatial SO(3) rotation group and a subgroup of this group relevant to
the symmetries in the presence of the fields is the molecular point group
$D_2=\{E,C_2^x, C_2^y,C_2^z\}$, where $E$ stands for the identity and $C_2^i$
represents a rotation of $\pi$  around the $i$-axis 
of the MFF, with $i=x,\quad y,\quad z$. 
The action of these operators on  the Euler angles are summarized in 
Table~\ref{table:symmetry}.  
The $D_2$-group has four irreducible
representations. 
The Wang states, defined as
\begin{align*}
\ket{JKMs}^w&=\cfrac{1}{\sqrt{2}}\left(\ket{JKM}+(-1)^s\ket{J-KM}\right),\;K>0
\\
 \ket{J0M0}^w&=\ket{J0M},\;K=0, 
\end{align*}
with $s=0$ and $1$, 
form the basis of these irreducible representations,
characterized by the parity of $J+s$ and $K$. The action of the elements of the
$D_2$-group is
$C_2^i\ket{JKMs}^w=(-1)^{\lambda_i}\ket{JKMs}^w$, with $i=x,y,z$
and  $\lambda_x=J+K+s$, $\lambda_y=J+s$, and  $\lambda_z=K$.
The states $\ket{JKM}$ are the eigenfunctions of the field-free symmetric top
rotor 
\begin{equation} 
\ket{JKM}=(-1)^{M-K}\sqrt{\frac{2J+1}{8\pi^2}}D_{-M,-K}^J(\Omega),
\label{eqn:sym_basis}
\end{equation}
with $D_{M,K}^J(\Omega)$ being the Wigner matrix elements~\cite{zare}, 
 $J$ the total angular momentum, and $K$ and $M$ the projections of
$\mathbf{J}$ on the MFF $z$-axis and on the LFF $Z$-axis,
respectively. To be self-contained, the definition and main properties of the Wigner matrix
elements are in Appendix \ref{sec:matrix}.

\begin{table}[t]  
  \begin{tabular}{cccc}
    \hline
    \hline
    &&Transformations&\\
    \hline
    Operation\;&$\phi$ &$\theta$ \;&$\chi$\\
    \hline
    $C_2^z$\,&$\phi\rightarrow\phi$\,&$\theta\rightarrow\theta$\,&$\chi\rightarrow\chi-\pi$\\
    $C_2^y$\,&$\phi\rightarrow\phi$\,&$\theta\rightarrow\pi-\theta$\,&$\chi\rightarrow-\chi$\\
    $C_2^x$\,&$\phi\rightarrow\phi-\pi$\,&$\theta\rightarrow\pi-\theta$\,&$\chi\rightarrow\pi-\chi$\\
    $\sigma_{XZ}^\dagger$\,&$\phi\rightarrow2\pi-\phi$\,&$\theta\rightarrow\theta$\,&$\chi\rightarrow
    \pi-\chi$\\
    $C_X(\pi)$\,&$\phi\rightarrow2\pi-\phi$\,&$\theta\rightarrow\pi-\theta$\,&$\chi\rightarrow\pi+\chi$\\
    $C^\alpha_{\perp Z}(\pi)$\,&$\phi\rightarrow2\alpha-\phi$\,&$\theta\rightarrow\pi-\theta$\,&$\chi\rightarrow\chi+\pi$\\
    $C_Z(\delta)$\,&$\phi\rightarrow\phi+\delta$\,&$\theta\rightarrow \theta$\,&$\chi\rightarrow\chi$\\
    \hline
    \hline
  \end{tabular}
\caption{Action of the symmetry operations on the Euler angles. 
$^\dagger$The reflection can not be represented only by a rotation and the
  operation $y\rightarrow-y$ should be performed as well.}
\label{table:symmetry}  
\end{table}

For a field-free asymmetric top rotor, $J$ and $M$ are good quantum
numbers, whereas, in contrast to a symmetric rotor, $K$ is not well defined. 
For each $M$-value, there are four irreducible representations that depend on
the parity of $J+s$ and $K$. The eigenstates are degenerate with respect to $M$, and for a
certain $M$ and $J$, the corresponding eigenfunctions are  linear
combinations of Wang states  $\ket{JKMs}^w$ with different $K$ values.   

Since an external field defines a preferred direction in space, the 
symmetries of the corresponding Hamiltonian are reduced compared to the
field-free case. As a consequence, the total angular momentum  
$J$ is not a good quantum number, and only for certain field configurations 
$M$ remains as a good quantum number. 

In the presence of a nonresonant laser field linearly polarized along the
$Z$-axis,  
the symmetry operations of the Hamiltonian are the $D_2$ point
group, a rotation of an arbitrary angle $\delta$ around the $Z$-axis
$C_Z(\delta)$,  a rotation of
$\pi$ around an axis perpendicular to the $Z$-axis tilted at an angle
$\alpha$ with respect to the
$X$-axis $C^\alpha_{\perp Z}(\pi)$, 
and the reflection in any plane including the $Z$-axis (this reflection is
equivalent to first applying a twofold rotation around any axis in the
$XY$-plane followed by the action of the operator $C_2^x$ or 
$C_2^y$~\cite{kanya:pra70}).  Only $M$ remains a
good quantum number, and the levels with  $\pm M$ are
degenerate. 
Thus, we have eight irreducible representations for
each $|M|>0$, but due to the reflection at the plane 
there is a twofold degeneracy and they can be effectively reduced 
to four representations being characterized by the parity of $K$ and $J+s$, as 
in the field-free Hamiltonian. For $M=0$, there are eight irreducible
representations labeled by the parities of $J$, $K$  and $s$. 

When an asymmetric top rotor is exposed to a static electric field (parallel
to the 
$Z$-axis) or to both fields in the parallel configuration, i.e., $\beta=0$, the
symmetry operations are 
$C_2^z$ from the $D_2$ point group, the rotation $C_Z(\delta)$,
and the reflection in any plane including the $Z$-axis. 
In these two cases, $M$ is still a good quantum number, and the states $M$ and
$-M$ are degenerate. For a certain $|M|$, the group has $4$ irreducible
representations characterized by the parity of $K$ and the parity of $s$. 
Those representations with the same parity of $K$ and  $M>0$ are degenerate in
energy.   
The symmetric top eigenfunctions \eqref{eqn:sym_basis}, with defined parity of
$K$, form a basis of this irreducible representation. 
For the $M=0$ case, these four representations are not energetically
degenerate.

For non parallel fields, $M$ ceases to be a good quantum number.
If the two fields are perpendicular, i.e., the electric
field is parallel to the $X$-axis and  $\beta=\pi/2$, the Hamiltonian
commutes with only three symmetry operations: 
$C_2^z$, a rotation of $\pi$ around the
$X$-axis $C_X(\pi)$, and the reflection $\sigma_{XZ}$ on the $XZ$-plane 
where the fields are contained, as well as their combinations. 
Using the field-free symmetric rotor wave functions, we
can construct a basis for  
$\sigma_{XZ}$, 
%\begin{align*}
% \ket{JKMq}^\sigma&=\frac{1}{\sqrt{2}}(\ket{JKM}+(-1)^q\ket{J-K-M}),\\
%&\quad \text{with}\quad M\; \text{and/or}\;K\ne0,\\ 
%\ket{J000}^\sigma&=\ket{J00},\\
%&\quad \text{with}\quad M=K=0, 
%\end{align*}
\begin{equation*}
 \ket{JKMq}^\sigma=\frac{1}{\sqrt{2}}(\ket{JKM}+(-1)^q\ket{J-K-M})
\end{equation*}
with $M\; \text{and/or}\;K\ne0$, and 
\begin{equation*}
\ket{J000}^\sigma=\ket{J00}
\end{equation*}
with $M=K=0$, 
where $q=0$ and $1$, and 
$\sigma_{XZ}\ket{JKMq}^\sigma=(-1)^{M+K+q} \ket{JKMq}^\sigma$; 
and one for $C_X(\pi)$ 
\begin{equation*}
\ket{JKMp}^X=\frac{1}{\sqrt{2}}(\ket{JKM}+(-1)^p\ket{JK-M}),
\end{equation*}
with $M\ne0$, and 
\begin{equation*}
\ket{JK00}^X=\ket{JK0},
\end{equation*}
with $M=0$, 
where $p=0$ and $1$, and 
 satisfying $C_X(\pi)\ket{JKMp}^X=(-1)^{J+p}\ket{JKMp}^X$. The group of
all the symmetry operators, $C_2^z$, $\sigma_{XZ}$ 
and $C_X(\pi)$, has eight different irreducible 
representations according to the parity
of $M+K+q$,  $J+p$ and $K$. 
A basis of these irreducible representations is
\begin{eqnarray*}
\ket{JKMqp}_{\frac{\pi}{2}}&=&\frac{1}{2}
(\ket{JKM}+(-1)^q\ket{J-K-M}\\
&&+(-1)^p\ket{JK-M}+(-1)^{p+q}\ket{J-KM}),
\end{eqnarray*}
with $ M\ne0$ and $K\ne0$, 
\begin{equation*}
  \ket{J0M0p}_{\frac{\pi}{2}}=\frac{1}{\sqrt{2}}(\ket{J0M}+(-1)^p\ket{J0-M}),
\end{equation*}
with $M\ne0$ and $K=0$ 
\begin{equation*}
  \ket{JK0q0}_{\frac{\pi}{2}}=\frac{1}{\sqrt{2}}(\ket{JK0}+(-1)^q\ket{J-K0}),
\end{equation*}
with $M=0$  and $K\ne0$ and 
\begin{equation*}
 \ket{J0000}_{\frac{\pi}{2}}=\ket{J00},
\end{equation*}
with $K=M=0$, 
where  $q=0$ and $1$,  $p=0$ and $1$, and
the parity of $M+K+q$, $K$ and $J+p$ are preserved. 

Finally, when the fields form an angle $0<\beta<\pi/2$, the
Hamiltonian is invariant under the reflection $\sigma_{XZ}$ and the rotation
$C_2^z$.  We have, therefore, four irreducible representations depending on
the parity of $M+K+q$ and of $K$, and  the corresponding 
basis is $\{ \ket{JKMq}^\sigma\}$.

For an asymmetric top rotor exposed to any of these field configurations, the  
field-dressed spectrum exhibits many avoided crossings between energetically
adjacent states of the same symmetry. 
When the spectrum is analyzed as the strength of one of these fields is varied
or the angle between them, these
avoided crossings should be distinguished from the real crossings taking place
between levels of
different symmetry. Hence, we solve the Schr\"odinger equation, by expanding
the rotational wave function in a 
basis that respects the symmetries of the corresponding  Hamiltonian.
As a consequence, the coefficients of these expansions fulfill the properties
of the basis vectors  of the corresponding irreducible
representation.
For computational reasons, we have cut the (in principle) infinite serie to a 
finite one including only those functions with $J\le J_\text{max}$, and
for a certain $J$ all $(2J+1)$-values of $K$, and, analogously, for $M$ in the
case $0<\beta\le\pi/2$. 
The size of the Hamiltonian matrix increases as $J_{\text{max}}^3$ and 
 $J_{\text{max}}^2$
for the $0<\beta\le\pi/2$ and $\beta=0$ configurations, respectively. 
In this study, we have used $J_\text{max}=24$, and the convergence is reached
for the states analyzed here. 
Several matrix elements are presented in Appendix \ref{sec:matrix}.  

The field-free states are labeled by  the notation $J_{K_a,K_c}M$, where $K_a$
and $K_c$ are the values of $K$ on the limiting symmetric top rotor 
prolate and oblate cases, respectively~\cite{king_jcp11}. 
For reasons of addressability, we use this notation for the field-dressed
states, even if $J$ and/or $M$ are not good quantum numbers. 
Thus, $J_{K_aK_c}M$ refers to the
level that is adiabatically connected as $I$, $E_S$  and/or $\beta$ 
are modified with the
field-free state  $J_{K_aK_c}M$. The irreducible representation to which the
states belong is also indicated. 
Analogously to a symmetric top molecule exposed to combined 
fields~\cite{hartelt_jcp128}, the final labels of the states 
depend on the path followed on the parameters to reach a certain
field configuration, i.e., monodromy is observed.  
Since each interaction breaks different symmetries of the field-free
Hamiltonian, the  
order the fields are turned on determines the evolution of the field-dressed 
states. The complexity of the spectrum is characterized by the
amount of genuine and avoided crossings among the states, 
the symmetry of the two levels determine the type of crossing that they  may suffer 
as one of the field parameters ($E_S$, $I$ or $\beta$) is varied, and, therefore, if 
the corresponding labels may or not be  interchanged.

\section{Results}
\label{sec:results}

\begin{figure}[h]
  \centering
  \includegraphics[scale=0.35]{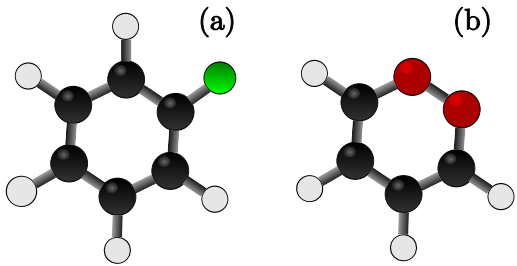}
  \caption{Structure of the (a) fluorobenzene and (b) pyridazine
molecules.}
  \label{fig:both_molecules}
\end{figure}

In this section, we illustrate the impact of the external fields
on two asymmetric top molecules:
fluorobenzene (C$_6$H$_5$F) and pyridazine (C$_4$H$_4$N$_2$).
Their data are summarized in Table~\ref{tab:fb_pyr}, according to 
Refs.~\cite{bak:jcp26,miller:jamchemsoc112,kowalewski:jcp31,innes:j_mol_spect_132_492,hinchliffe:jmst_304}
and their structure is shown in Fig. \ref{fig:both_molecules}. 
They are characterized by a different degree of asymmetry, the 
fluorobenzene is intermediate-prolate with $\kappa=-0.5879$, and the
pyridazine is near-oblate with $\kappa=0.8824$. 
The permanent dipole moment of  pyridazine is around 2.5 times larger than 
in fluorobenzene. 
The asymmetry of the polarizability tensor is very similar for both systems, 
for fluorobenzene, $\alpha^{zx}=4.298 \text{\AA}^3$ and 
$\alpha^{yx}=3.848 \text{\AA}^3$, 
whereas for pyridazine $\alpha^{zx}=4.51 \text{\AA}^3$ and 
$\alpha^{yx}=4.45 \text{\AA}^3$. 
Since the rotational constants of pyridazine are larger than for
fluorobenzene, for the same laser intensity, a weaker 
impact on the former should be expected.
In the following, we carry out a study of the spectrum of these two systems 
as the parameters that characterize the field configurations, $E_S$, $I$ or
$\beta$, are modified. 
For the sake of simplicity, we analyze the energy, the
orientation, the alignment and hybridization of the states which belong to the
representation with $K$ and $M+K+q$ even (if $\beta\ne\pi/2$),
and $J+p$  even (if $\beta=\pi/2$). 
Note that they represent well the main physical features observed in
the overall spectrum, and similar  behavior and properties are therefore obtained 
for the other representations.

\begin{table}[h]
  \centering
  \begin{tabular}{ccc}
    \hline
    \hline
    & Fluorobenzene\; & Pyridazine\\
    \hline
    $B_x$ $(\text{MHz})$\hspace{.6 cm} & 1716.916 & 3055.485\\
    $B_y$ $(\text{MHz})$\hspace{.6 cm} & 2570.624 & 6048.613\\
    $B_z$ $(\text{MHz})$\hspace{.6 cm} &5663.72 & 6235.680\\
    $\kappa$\hspace{.6 cm} & -0.5879 & 0.8824\\
    $\mu_z$     $(D)$\hspace{.6 cm} & 1.66 & 4.14\\
    $\alpha_{xx}$ $(\text{\AA}^3)$\hspace{.6 cm} & 7.141 & 5.84\\
    $\alpha_{yy}$ $(\text{\AA}^3)$\hspace{.6 cm} & 10.89 & 10.29\\
    $\alpha_{zz}$ $(\text{\AA}^3)$\hspace{.6 cm} & 11.439 & 10.35\\
    \hline
    \hline
  \end{tabular}
  \caption{Relevant data for 
fluorobenzene~\cite{bak:jcp26,miller:jamchemsoc112,kowalewski:jcp31} and
pyridazine~\cite{innes:j_mol_spect_132_492,hinchliffe:jmst_304}.} 
  \label{tab:fb_pyr}
\end{table}

\subsection{Impact of a linearly polarized laser field}
\label{sec:laser}

\begin{figure}
  \centering
\includegraphics[scale=.8,angle=0]{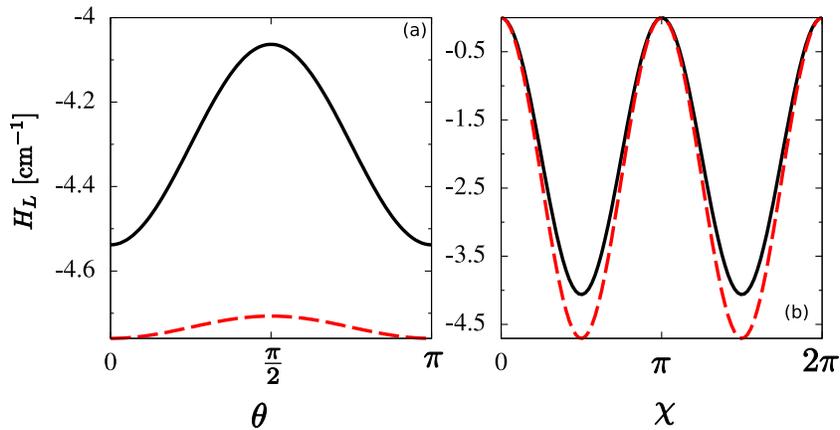}
\caption{Laser interaction term  $H_L$, see eq.  
\eqref{eqn:hamiltonian_laser},  in units of cm$^{-1}$ 
with $I=10^{11}$ Wcm$^{-2}$, for $\chi=\pi/2$ (a) and $\theta=\pi/2$ (b)
  for fluorobenzene (solid) and pyridazine (dash).} 
\label{fig:laser_interaction_alone}
\end{figure}
The dynamics of the molecule in the presence of a linearly polarized laser
field depends strongly on the anisotropy of the polarizability tensor.
The interaction  with the laser (\ref{eqn:hamiltonian_laser}) exhibits
several critical points. 
The minimum value $H_L=- I\alpha^{zx}/2\epsilon_0c$ is reached
for $\theta=0$ or $\pi$ and any value of $\chi$.
The interaction achieves three maxima at
$\theta=\pi/2$ and $\chi=0,\pi$ or $2\pi$ satisfying that $H_L=0$,  and
two saddle points at  $\theta=\pi/2$ and $\chi=\pi/2$ or $3\pi/2$ 
with $H_L=-I\alpha^{yx}/2\epsilon_0c$.
All of them are shown in  
Figs. \ref{fig:laser_interaction_alone}(a) and (b), where
$H_L$  is plotted for $\chi=\pi/2$ and $0\le\theta\le 2\pi$, and for
$\theta=\pi/2$ and $0\le\chi\le2\pi$, respectively. 
The main difference between the two systems is that 
the values for $H_L$ at the saddle point $\theta=\pi/2$ and
$\chi=\pi/2$ is smaller for pyridazine compared to fluorobenzene,
and the fact that the shape of $H_L$ as a function of $\theta$ and for fixed $\chi$ is
significantly flatter for the former. Taking into account that the rotational
constants are much larger for pyradizine compared to fluorobenzene we observe that 
for the same laser intensity $I$, the pyridazine wave function 
is more widespread with respect to $\theta$ and $\chi$ 
than for fluorobenzene.
While the minima are responsible for the molecular alignment,
the maxima or saddle points correspond to an "antialigned" wave function, 
that is the dipole moment points perpendicular to the field direction.

\begin{figure}
  \centering
\includegraphics[scale=.8,angle=0]{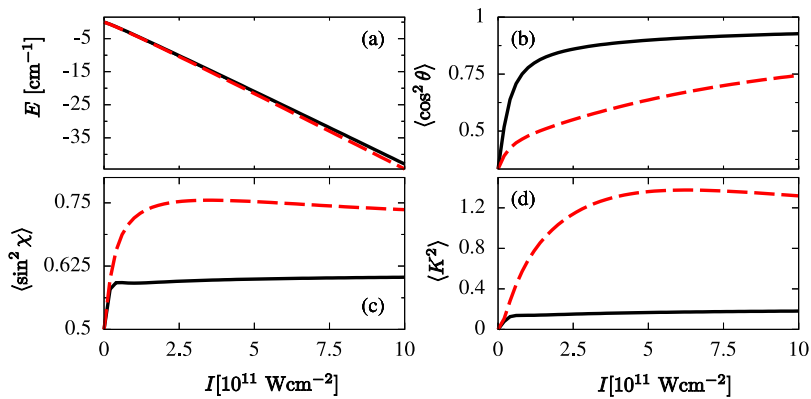}
\caption{Ground state energy (a), $\langle\cos^2\theta\rangle$ (b),
  $\langle\sin^2\chi\rangle$  (c),  and $\langle K^2\rangle$ (d),  
as a function of the
  intensity of a linearly polarized laser field, for the 
  fluorobenzene (solid) and pyridazine (dash) molecules.} 
\label{fig:laser_groundstate} 
\end{figure}
To get a better physical insight into this interaction, we present 
its impact on the ground state energy, and the expectation values
$\expected{\cos^2\theta}$, $\expected{\sin^2\chi}$, and  $\expected{K^2}$ in
Figs.~\ref{fig:laser_groundstate}(a), (b), (c), and (d), respectively. 
While, for both molecules, the energy as a function of $I$ shows a  
similar decreasing behavior, their values being indistinguishable on the
scale of Fig.~\ref{fig:laser_groundstate}(a), significant differences are observed for the  
other quantities.
In the very strong laser field regime, the probability density
 of  the fluorobenzene ground state tends to concentrate around the minima, and 
one should expect that $\expected{\cos^2\theta}\rightarrow 1$, and
$\expected{\sin^2\chi}\rightarrow 0.5$ for very large  intensities.
This last relation holds because the $\chi$-coordinate does not play any role
in the absolute minima of $H_L$, and one concludes that the probability density should be uniformly
distributed with respect to $\chi$.
Numerically we obtain that
for the fluorobenzene ground state, $\expected{\cos^2\theta}$ increases until
$0.76$ for $I=10^{11}$ Wcm$^{-2}$, decreasing thereafter, 
 and $\expected{\sin^2\chi}$ reaches a plateau
with a constant value $0.58$ for $I\ge4.2\cdot 10^{10}$ Wcm$^{-2}$.
The barrier height of $H_L$ as a function of $\theta$ for a
certain value of $\chi$, see Fig.~\ref{fig:laser_interaction_alone}(a), is
around $9.2$ times smaller for pyridazine than for  fluorobenzene, 
and for the former, the  rotational constants are larger, whereas  the polarizability anisotropies
$\alpha^{zx}$ and $\alpha^{yx}$
are of the same order for both molecules.
Hence, compared to fluorobenzene, the pyridazine ground state wave function
should be spatially stronger delocalize with respect to $\theta$,
and, therefore,  less aligned 
for the same laser intensity. We obtain here
$\expected{\cos^2\theta}=0.49$  for $I=10^{11}$ Wcm$^{-2}$.
As a consequence, the spreading
of the wave function for $\chi$ is not observed in  pyridazine: $\expected{\sin^2\chi}$  shows a maximum of
$0.75$ for $I=3 \cdot 10^{11}$ Wcm$^{-2}$ and
slightly decreases with further increasing $I$. 
For the field-free ground state, we have $\expected{K^2}=0$, and
as $I$ is increased $\expected{K^2}$ follows a similar evolution as 
$\expected{\sin^2\chi}$.
For fluorobenzene, $\expected{K^2}$ achieves the value $0.14$ for
$I\approx 7.2\times 10^{10}$ Wcm$^{-2}$, followed by a plateau-like behavior 
around  $\expected{K^2}\approx0.18$ 
for larger intensities for fluorobenzene, while it keeps an
increasing trend up to
$1.38$ for $I=5.9\times 10^{11}$ Wcm$^{-2}$ and decreasing smoothly afterwards  
for pyridazine. 
Let us emphasize that, for a field-free asymmetric rotor, $K$ is not a good
quantum number,  and an eigenstate already shows
a certain amount of $K$-mixing, but in the strong laser field regime, the second
term of the laser interaction \eqref{eqn:hamiltonian_laser}
should impact and enhance this $K$-mixing.

\subsection{Constant static electric field and increasing laser intensity} 
\label{sec:increasing_laser}
In the presence of an additional static field, the interaction is given by
$H_S+H_L$, see eqs. \eqref{eqn:hamiltonian_static} and
\eqref{eqn:hamiltonian_laser}, and the dynamics is significantly more
complicated. The amount of extremal points of this potential and their
character strongly depend on the field parameters as well as on the molecular
polarizability and permanent dipole moment.

\begin{figure}[t]
\includegraphics[scale=.8,angle=0]{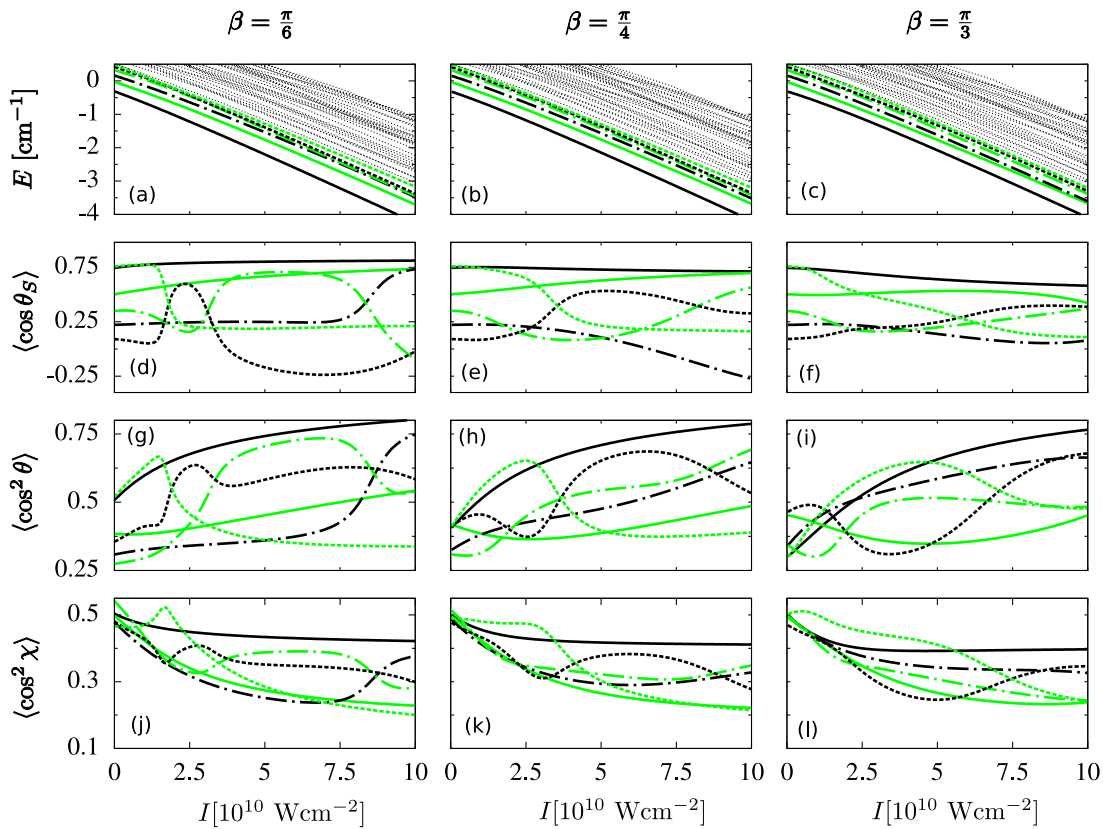}
\caption{(a)-(c) Energies and the expectation values (d)-(f)
  $\langle\cos\theta_S\rangle$, (g)-(i) $\langle\cos^2\theta\rangle$ and
  (j)-(l) $\langle \cos^2\chi\rangle$ for a constant field $E_S=20$ kVcm$^{-1}$ as a
  function of the intensity of the laser field for
  $\beta=\pi/6,\;\pi/4\; \text{and}\; \pi/3$ for the
  first states with both $M+q+K$ and $K$ even for fluorobenzene. The states are
  $0_{00}0$ (solid black), $1_{01}1$ (solid green), $1_{01}0$ (dash dotted
    black), $2_{02}2$ (dash dotted   green), $2_{02}1$ (dash
  black) and $3_{03}2$ (dash green). The spectrum (a, b, c)
  contains also highly excited states (very thin lines).} 
\label{fig:c6h5f_Static_20_laser_1_10_10_mq_even_k_even}
\end{figure}
\begin{figure}[bh]
  \centering
\includegraphics[scale=.8,angle=0]{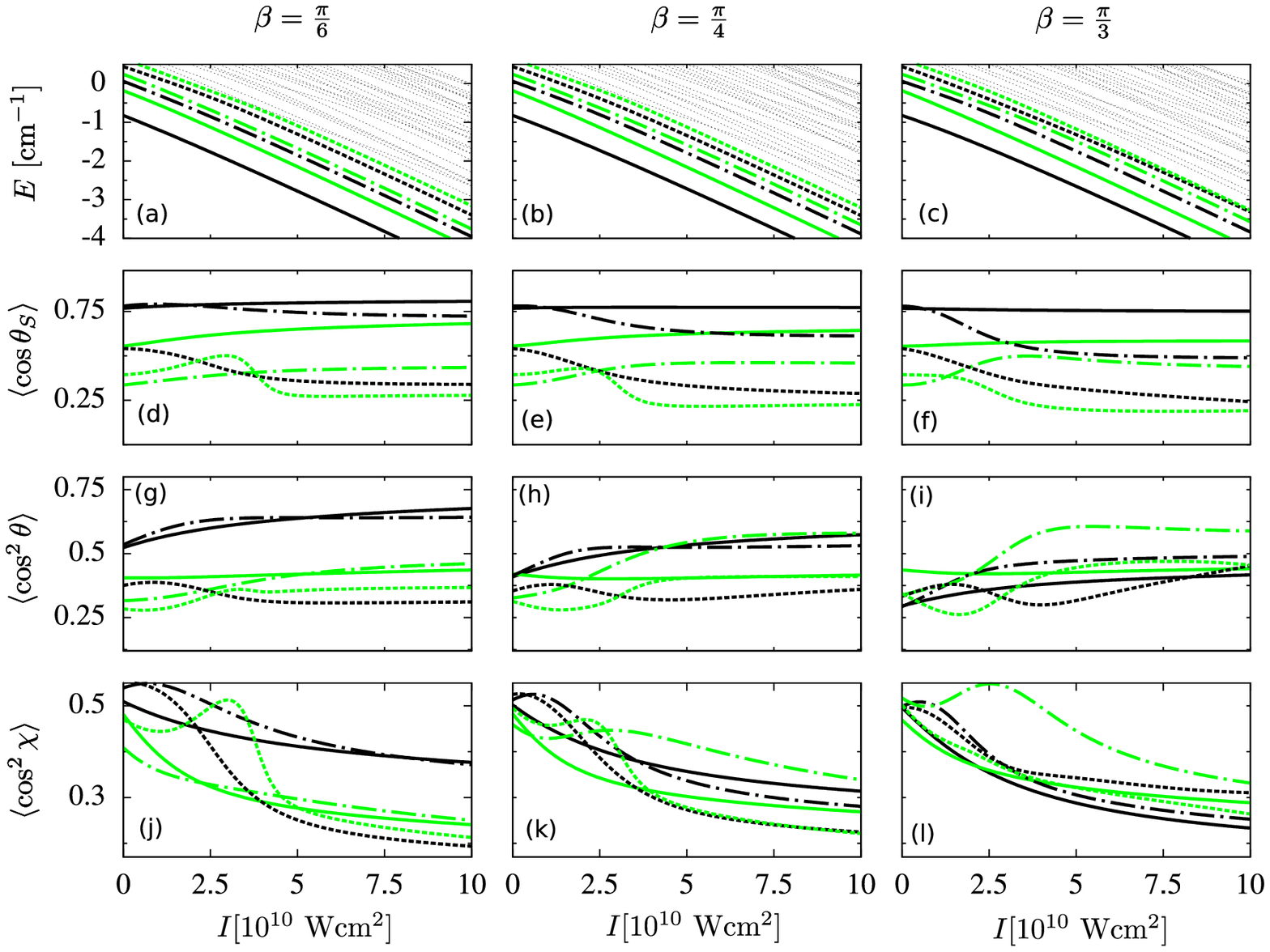}
\caption{(a)-(c) Energies and the expectation values (d)-(f)
  $\langle\cos\theta_S\rangle$, (g)-(i) $\langle\cos^2\theta\rangle$ and
  (j)-(l) $\langle \cos^2\chi\rangle$ for a constant field $E_S=20$ kVcm$^{-1}$ as a
  function of the intensity of the laser field for
  $\beta=\pi/6,\;\pi/4\; \text{and}\;\pi/3$ for the
  first states with both $M+q+K$ and $K$ even for pyridazine. The states are
  $0_{00}0$ (solid black), $1_{01}1$ (solid green), $2_{02}2$ (dash dotted   black), $1_{01}0$ (dash dotted   green), $2_{02}1$ (dash
  black) and $2_{21}2$ (dash green). The spectrum (a, b, c)
  contains also highly excited states (very thin lines).} 
\label{fig:pirazine_Static_20_laser_1_10_10_mq_even_k_even}
\end{figure}

For fluorobenzene and pyridazine, we show in
Figs. \ref{fig:c6h5f_Static_20_laser_1_10_10_mq_even_k_even}
and   \ref{fig:pirazine_Static_20_laser_1_10_10_mq_even_k_even} 
the dependence of the energies  (panels  (a), (b), and  (c)), 
expectation values $\langle\cos\theta_S\rangle$ (panels (d), (e), and (f)), 
$\langle\cos^2\theta\rangle$ (panels  (g), (h), and (i)), and
$\langle\cos^2\chi\rangle$ (panels (j), (k), and (l)),  
as functions of the laser intensity for a constant electric field
$E_S=20\;\text{kVcm}^{-1}$, and $\beta=\pi/6, \,\pi/4$ and $\pi/3$, respectively. 
For the sake of simplicity and without loosing generality, we restrict our
analysis to the energetically lowest-lying six states for the irreducible
representations with both
$K$  and $M+K+q$ being even, 
which are the levels 
$J_{K_a,K_c}M=0_{00}0$, $1_{01}0$, $1_{01}1$, $2_{02}1$, $2_{02}2$ 
and $3_{03}2$ for  fluorobenzene, and
$J_{K_a,K_c}M=0_{00}0$, $1_{01}0$, $1_{01}1$, 
$2_{02}1$, $2_{02}2$, and $2_{21}2$, for  pyridazine.
To illustrate the complexity of the spectrum, we have included in the energy
panels highly excited levels with the same symmetry (very thin lines). 
The adiabatic following has been done by increasing first 
the strength of the static electric
field up to $20$ kVcm$^{-1}$, the static field being tilted by an angle $\beta$ with
respect to the $Z$-axis. This is done for $I=0$ and yields the labeling of the
states in the presence of the static field. Thereafter, the laser intensity is increased. 
For both molecular systems, these levels are high-field-seekers, and their energies 
decrease as $I$ is increased. For a given laser intensity, 
the lowering in energy (compared to the field-free value) increases with
decreasing angle between both
fields. Since all the states included in these figures possess the same symmetry,
we encounter exclusively avoided crossings of energetically adjacent states. We have assumed that the avoided crossings are 
traversed adiabatically as $I$ is increased (according to Landau-Zener
transition theory), and consequently the character of the involved states is interchanged.
These avoided crossings, which are not distinguishable on the energy scale panels (a)-(c) in Figs.
\ref{fig:c6h5f_Static_20_laser_1_10_10_mq_even_k_even} and
\ref{fig:pirazine_Static_20_laser_1_10_10_mq_even_k_even}, 
strongly affect the orientation and alignment features of these levels.

Let us start analyzing the results for fluorobenzene in 
Fig.~\ref{fig:c6h5f_Static_20_laser_1_10_10_mq_even_k_even}. 
The orientation of the corresponding wave functions is illustrated by the
expectation value 
$\langle\cos\theta_S\rangle$, with $\theta_S$ being the angle between the
static electric field and the molecular 
fixed $z$-axis, which coincides with the direction of the permanent dipole moment.
Only the states $0_{00}0$ and $3_{03}2$ present a
significant orientation  
with respect to the static electric field direction, that is reduced as  $\beta$ is increased. 
The ground state satisfies $\langle\cos\theta_S\rangle>0.70$ for the
three $\beta$ values, and  
it has a plateau like behavior, with a minor positive or negative slope as $I$ is increased.
The numerous avoided crossings have significant impact on the other levels
especially for $\beta=\pi/6$ and $\pi/4$, and  
the underlying states might evolve from a strongly oriented configuration into a weakly
oriented or antioriented one. 
As an example, the state $3_{03}2$ after suffering for 
$I\approx1.7 \cdot 10^{10}$ Wcm$^{-2}$ an avoided
crossing with the non-oriented $2_{02}1$ state looses its strong orientation. 
By further increasing $I$ the $2_{02}1$ level suffers another avoided
crossing, which provokes a  
local maximum in  $\langle\cos\theta_S\rangle$ for 
$I\approx2.4 \cdot 10^{10}$ Wcm$^{-2}$. 
Since for $\beta=\pi/3$ the width of all the avoided crossings is larger 
for all the levels,  the  orientation shows a smooth evolution  as $I$ is
varied. 
A nonresonant linearly polarized laser field provokes the alignment of the 
wave function along
the $Z$-axis, i.e., $\expected{\cos^2\theta}$ tends to increase and ultimately
approach the value $1$ as
$I$ is increased
(Fig.~\ref{fig:c6h5f_Static_20_laser_1_10_10_mq_even_k_even}(g, h, i)). This
process competes now with the orientation due to the
static field. Only the ground state alignment keeps  an increasing trend as
$I$ is enhanced, and for the given three configurations  
$\expected{\cos^2\theta}>0.75$ for $I=10^{11}$ Wcm$^{-2}$. 
An additional electric field at an angle $\beta=\pi/6$ or
$\pi/4$ favors the alignment and $\expected{\cos^2\theta}$ is larger 
than without static field, see Fig.~\ref{fig:laser_groundstate}(b), whereas as
the angle between the fields is augmented the values achieved for
$\expected{\cos^2\theta}$ come closer to those of 
Fig.~ \ref{fig:laser_groundstate}(b).
The level $1_{01}1$ does not achieve a large alignment, and for $\beta=\pi/4$, 
$\expected{\cos^2\theta}$ exhibits a broad well that for
$\beta=\pi/3$ is even wider since the coupling between the states
changes.
Around the avoided crossings, the wave function of the involved states 
alternate regions of significant alignment with other characterized by broad
distribution as the laser intensity is varied, e. g. see the aligment of the 
states $3_{03}2$
and $2_{02}1$ for $I\approx1.7\cdot 10^{10}$ and $3.5\cdot
10^{10}$ Wcm$^{-2}$ in
Fig.~\ref{fig:c6h5f_Static_20_laser_1_10_10_mq_even_k_even} (g) and (h),
respectively. The impact of the avoided crossings is also noticeable for
$\beta=\pi/3$. The behavior of $\expected{\cos^2\chi}$ strongly depends on the
considered level. Since the molecules exhibit a strong orientation with the
electric field, the contribution in $\chi$ should be to increase the term
$\expected{\sin^2\chi}$ which will give rise to a decrease of the energy. 
For the states $0_{00}0$ and $1_{01}1$, $\expected{\cos^2\chi}$
decreases as $I$ is increased, and for both states there exists some region
where $\expected{\cos^2\chi}$ keeps a smooth behavior. 
For the other states, this expectation value is also affected by the presence
of avoided crossings, and  $\expected{\cos^2\chi}$  alternates between
increasing and decreasing behavior as a function of the laser intensity.
Since the pyridazine possesses a larger permanent dipole moment than
fluorobenzene, the impact of the static field is larger, and also dominates
the dynamics, see
Fig.~\ref{fig:pirazine_Static_20_laser_1_10_10_mq_even_k_even}. 
For the ground state  and first excited one, $\expected{\cos\theta_S}$
is only weakly affected by the laser field, for the former 
$\expected{\cos\theta_S}\approx 0.75$ independently of $I$ and $\beta$, while
for the $1_{01}1$ level $\expected{\cos\theta_S}>0.5$. 
The other levels present a right-way orientation, that can be converted from a
strong to a mild one when an avoided crossing is encountered.
Regarding the alignment, for most of the states
 $\expected{\cos^2\theta}$ shows a smooth behavior as $I$ is enhanced for the
three values of $\beta$. The impact of the avoided crossings on this
expectation value is not very pronounced because 
most of the states show a weak alignment with
a similar value of $\expected{\cos^2\theta}$. 
Analogously to fluorobenzene, the ground state has a larger alignment for
$\beta=\pi/6$ and $\pi/4$ than in the absence of the static field, while for 
$\beta=\pi/3$ other states are found  with larger alignment.
For stronger fields, $\expected{\cos^2\chi}$ monotonically decreases as 
$I$ is enhanced for all the states, and the slope is more pronounced compared
to fluorobenzene.
  
\subsection{Constant laser intensity and increasing electric field strength} 
\label{sec:increasing_static}
\begin{figure}[b]
  \centering
\includegraphics[scale=.8,angle=0]{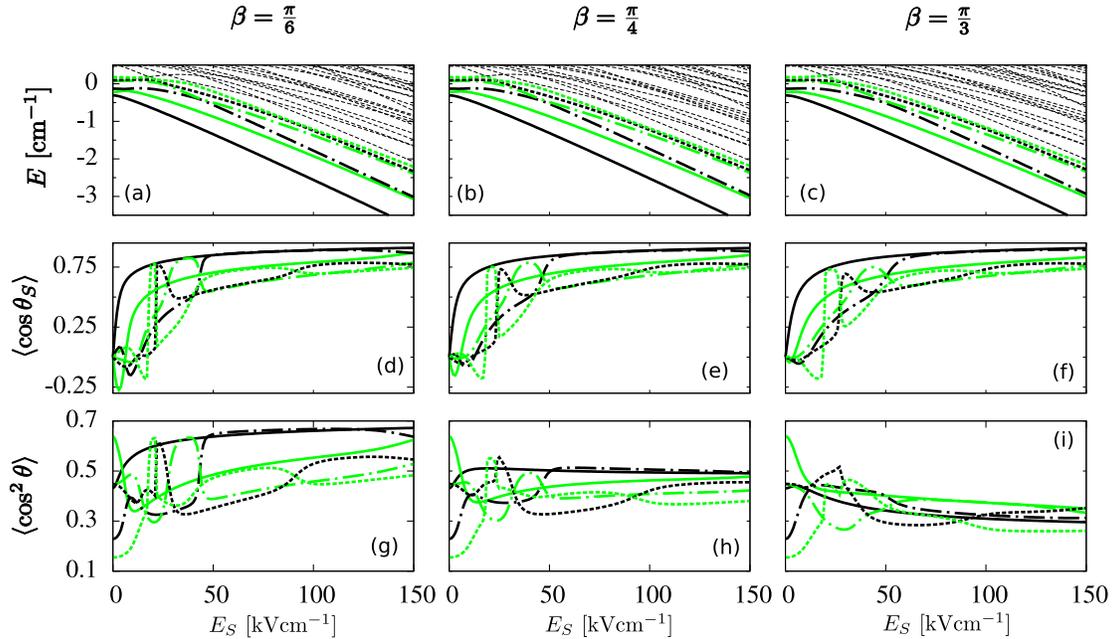}
\caption{(a)-(c) Energies and the expectation values (d)-(f)
  $\langle\cos\theta_S\rangle$, (g)-(i) $\langle\cos^2\theta\rangle$ for a constant $I=10^{10}$ Wcm$^{-2}$ as a
  function of the strength of the static field for
  $\beta=\pi/6,\; \pi/4\; \text{and}\; \pi/3$ for the
  energetically lowest states with both $M+q+K$ and $K$ even for fluorobenzene. The states are
  $0_{00}0$ (solid black), $1_{01}0$ (solid green), $1_{01}1$ (dash dotted
  black), $2_{02}0$ (dash dotted   green), $2_{02}1$ (dash black) and
  $2_{02}2$ (dash green). The spectrum (a, b, c)
  contains also highly excited states (very thin lines).} 
\label{fig:c6h5f_Static_0_200_laser_1_10_10_mq_even_k_even}
\end{figure}
\begin{figure}[bh]
  \centering
\includegraphics[scale=.8,angle=0]{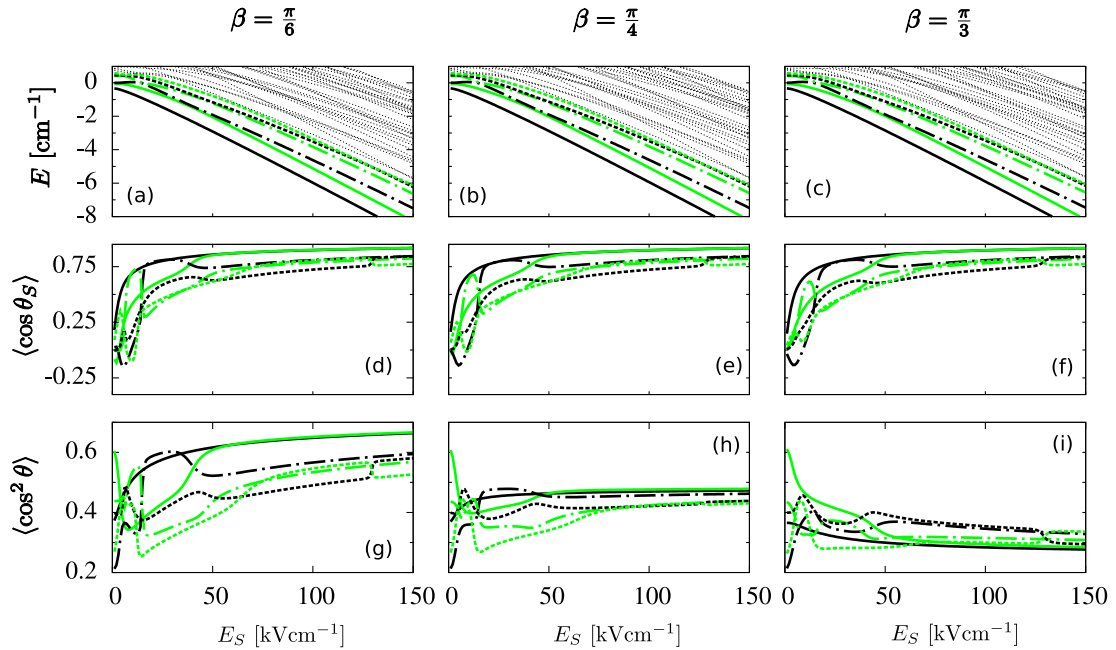}
\caption{Same as
  Fig.~\ref{fig:c6h5f_Static_0_200_laser_1_10_10_mq_even_k_even}
but for pyridazine.}
%(a)-(c) Energies and the expectation values (d)-(f)
%  $\langle\cos\theta_S\rangle$, (g)-(i) $\langle\cos^2\theta\rangle$ and
%  (j)-(l) $\langle \cos^2\chi\rangle$ for a constant $I=10^{10}$ Wcm$^{-2}$ as a
%  function of the strength of the static field for
%  $\beta=\pi/6,\; \pi/4\; \text{and}\; \pi/3$ for the
%  first states with $M+q+K$ even and $K$ even of pyridazine. The states are
%  $0_{00}0$ (solid black), $1_{01}0$ (solid green), $1_{01}1$ (dash
%  thin black), $2_{02}0$ (dash dotted   green), $2_{02}1$ (dash
%  thick black) and $2_{02}2$ (dash green). 
\label{fig:pirazine_Static_0_200_laser_1_10_10_mq_even_k_even}
\end{figure}

The pendular limit of an asymmetric top 
molecule in the presence of a strong electrostatic field was investigated 
by Kanya and Ohshima~\cite{kanya:pra70} using a power series expansion in $\mu
E_S$. Their analytical expression for the energy, which 
neglects the contribution of terms in powers equal or smaller than $(\mu
E_S)^{-1/2}$ 
(see eq. (26) in Ref.~\onlinecite{kanya:pra70}) allowed us for  a
straightforward comparison to our numerical calculations. 
For $E_S=100$ kVcm$^{-1}$, 
the energy and orientation cosine of the ground state
agree within $0.05\%$ and $0.025\%$ for fluorobenzene and $0.04\%$ and
$0.019\%$ pyridazine, respectively. 
Note that for highly excited states these relative errors increase.

For a constant laser field $I=10^{10}\;\text{Wcm}^{-2}$, we now investigate the impact
of increasing static field strength for three different configurations.  
Again, we consider the energetically lowest-lying six states with the
irreducible representation for
$K$  and $M+K+q$ being even.
These levels have been adiabatically followed as the laser intensity is raised
from $I=0$ to $10^{10}\;\text{Wcm}^{-2}$ for $E_S=0$, we
label them, and finally the electrostatic field is turned on forming an angle
$\beta$ and its strength is increased.  
Thus, for both molecules the levels are
$0_{00}0$, $1_{01}0$, $1_{01}1$, $2_{02}0$,  
$2_{02}1$, and $2_{02}2$.
For fluorobenzene and pyridazine,  we present 
in Figs. \ref{fig:c6h5f_Static_0_200_laser_1_10_10_mq_even_k_even} and
\ref{fig:pirazine_Static_0_200_laser_1_10_10_mq_even_k_even}, 
the evolution of the energies
(panels (a), (b), and (c)),
$\langle\cos\theta_S\rangle$
(panels (d), (e), and (f)),
and $\langle\cos^2\theta\rangle$ 
(panels (g), (h), and (i)),
as $E_S$ is increased, for $I=10^{10}\;\text{Wcm}^{-2}$, 
and $\beta=\pi/6,\pi/4$ and $\pi/3$, respectively.
We remark that for such a weak laser intensity $I=10^{10}$ Wcm$^{-2}$ a 
static electric field
of $E_S\ge 17.08$ and $7.26$ kVcm${^{-1}}$ for
fluorobenzene and pyridazine, respectively, 
provides the larger contribution to the external
field Hamiltonian $H_S+H_L$.
If the interaction with the static field is
dominant and much larger than the laser one, i.e.,
$E_s\mu\gg  I\alpha^{ix}/2\epsilon_0c$ with $i=y$ or $z$, the absolute minima of
the potential $H_S+H_L$ are at $(\phi,\beta,\chi)$, with $\phi\to 0 , 2\pi$ and
$\chi\to \pi/2, 3\pi/2$. 
Thus, the wave function will be oriented toward the electric field direction
with $\langle\cos^2\theta\rangle\to \cos^2\beta $ and 
$\langle\cos\theta_S\rangle\to 1$.  
However, these extremal points do not provoke any effect on the $\chi$
coordinate, because in the strong static field regime, the variation of
$H_S+H_L$ as a function of $\chi$ represents a very shallow
minimum and does not give rise to a localization of the wave function.

In the weak field regime there are, for both molecules, several levels in
addition to the ground state that are high-field seekers, and remaining levels
are for low values of $E_S$ low-field seekers.  
These levels present a mild wrong-way orientation with 
$\expected{\cos\theta_S}<0$. 
Since in this regime the slope of the variation of the energy with $E_S$
might be positive or negative, this favors the presence of sharp avoided
crossings. As a consequence, the orientation of a pair of levels 
involved in an avoided crossings, i.e., $\expected{\cos\theta_S}$,
suffers drastic variations over tiny ranges of the field strength.  
As the field strength is increased, we encounter the pendular regime: 
all the states are high-field-seekers, and they are strongly oriented along
the static field direction. Indeed, for $E_S=100$ kVcm$^{-1}$, we have that
$\expected{\cos\theta_S}>0.60$ and $0.70$ for the considered fluorobenzene and
pyridazine levels, respectively. In this regime, we still encounter avoided
crossings but they are much wider.
Due to the competition between both fields, these states do not achieve a
significant alignment, see panels (g)-(i) in 
Fig.~\ref{fig:c6h5f_Static_0_200_laser_1_10_10_mq_even_k_even} and
\ref{fig:pirazine_Static_0_200_laser_1_10_10_mq_even_k_even}.
We observe that $\expected{\cos^2\theta}$ approaches $\cos^2\beta$ in the
strong field regime.

\subsection{Orientation and 2-D alignment by means of perpendicular fields}
\label{sec:one-d-proba}

\begin{figure}
  \centering
  \includegraphics[scale=.8,angle=0]{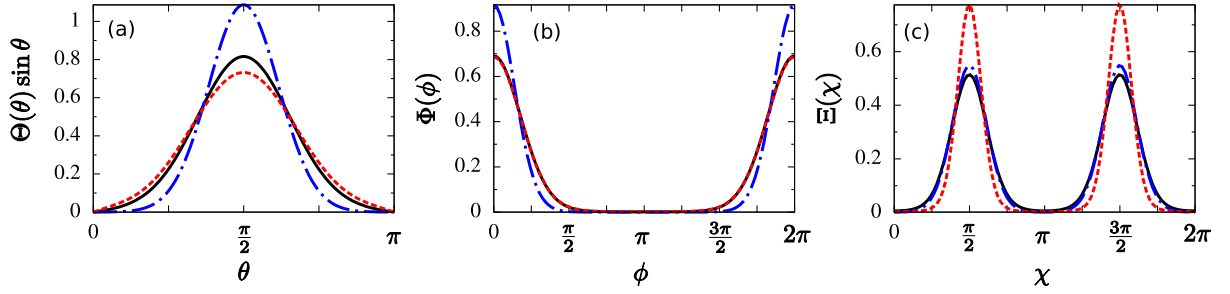}
  \caption{One dimensional probability density distribution in each Euler angle for the
    ground state of pyridazine for $E_S=20$ kVcm$^{-1}$ and
    $I=10^{11}$ Wcm$^{-2}$ (solid), $E_S=50$ kVcm$^{-1}$ and $I=10^{11}$
    Wcm$^{-2}$ (dash dotted), $E_S=20$ kVcm$^{-1}$ and $I=5\cdot
    10^{11}$
 Wcm$^{-2}$ (dash).}
  \label{fig:density_together}
\end{figure}

External fields provide a tool to control the molecular dynamics: specifically
it has been shown that  an elliptically polarized laser allows for
 \emph{3-D alignment} of asymmetric molecules, i.e., the 
system is aligned in all spatial 
directions~\cite{stapelfeldt:rev_mod_phys_75_543, larsen:phys_rev_lett_85_2470,
rouze:phys_rev_a_77_043412, artamonov:phys_rev_a_82_023413}. 
We show here that the combination of an electrostatic field with the linearly
polarized laser gives rise to orientation in one direction and alignment in
the other two. 
We hereby focus on the case $\beta=\pi/2$. In the strong static field regime, 
we have $\theta\rightarrow \pi/2$ and $\phi\rightarrow 0$,
whereas, in order to decrease the
energy the term $\sin^2\chi$ in the laser Hamiltonian \eqref{eqn:hamiltonian_laser}
should increase and approach one, i.e., $\chi\to \pi/2, 3\pi/2$. 
Then, the molecule will be fixed in space, and the wave function should be
concentrated in the proximity of 
$\theta\sim \pi/2$, $\chi\sim \pi/2,\; 3\pi/2$ and
$\phi\sim 0,\, 2\pi$. 
To demonstrate this behavior we have computed the 1D probability density
distribution in each one of the three Euler angles by integrating the square
of the wave function in the other two angles. 
In Fig.~\ref{fig:density_together}(a), (b), and (c), these 
probability density distributions 
$\Theta(\theta)\sin\theta$, $\Phi(\phi)$ and $\Xi(\chi)$ are plotted as a
function of $\theta$,
$\phi$ and $\chi$, respectively, for the ground state of pyridazine interacting 
with the orthogonal fields for different strengths. 
The probability density distributions for $E_S=20$ kVcm$^{-1}$ and 
$I=10^{11}$ Wcm$^{-2}$ show the expected behavior, and 
this state presents a significant orientation along the 
$X$-axis of the LFF with 
$\expected{\cos\theta_S}=0.749$ and $\expected{\cos^2\theta}=0.191$, 
whereas the alignment in the other two Euler angles is also pronounced 
and we get $\expected{\cos^2\chi}=0.143$ and 
$\expected{\cos^2\phi}=0.661$. 
A larger orientation is achieved if the static field strength is enhanced to 
$E_S=50$ kVcm$^{-1}$ keeping the same laser intensity:
$\expected{\cos\theta_S}=0.849$ and 
$\expected{\cos^2\theta}=0.118$, 
for the distribution in $\chi$ and $\phi$ and we find 
$\expected{\cos^2\chi}=0.122$  and 
$\expected{\cos\phi}=0.723$, respectively. 
An opposite effect has the enhancement of the laser intensity to 
$I=5\cdot 10^{11}$ Wcm$^{-2}$ and keeping  $E_S=20$ kVcm$^{-1}$. The
orientation in $\theta$ is slightly reduced $\expected{\cos\theta_S}=0.729$ and
$\expected{\cos^2\theta}=0.222$, but $\expected{\cos^2\phi}$ increases only
to $0.682$,
that is, the laser does not affect the azimuthal angle and in 
Fig.~\ref{fig:density_together}(b) the distribution $\Phi(\phi)$ is
indistinguishable compared to the corresponding one 
for $E_S=20$ kVcm$^{-1}$ and  $I=10^{11}$ Wcm$^{-2}$.
The alignment with respect to $\chi$ becomes stronger $\expected{\cos^2\chi}=0.075$.
Of course, one should keep in mind that this effect is very 
sensitive to the fields strengths and their configuration, as well as and the molecular parameters, i. e., the
permanent dipole moment, polarizability anisotropies and rotational
constants. For fluorobenzene, a similar phenomenon could be found, but
the wave function shows a slightly less pronounced orientation and 
alignment. If the fields are not perpendicular, the competition between both
interactions will reduce the alignment achieved in the $\phi$ and $\chi$
angles.

\subsection{Influence of the inclination of the fields}
\label{sec:inclination}
\begin{figure}[t]
 \centering
\includegraphics[scale=.8,angle=0]{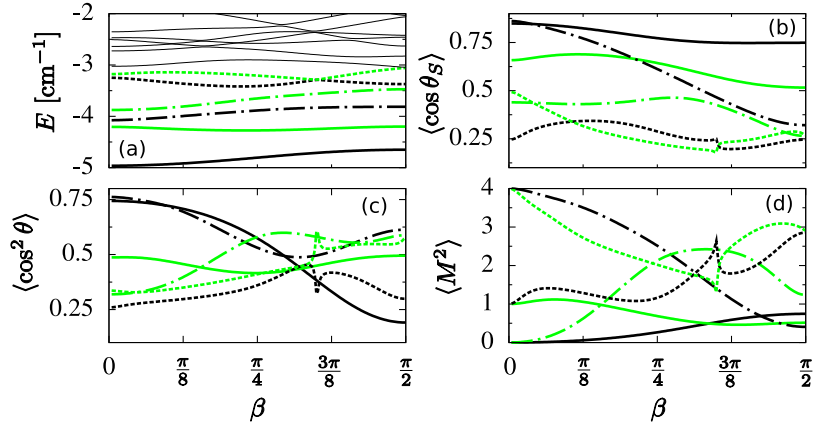}
\caption{(a) Energies and expectation values (b)
  $\langle\cos\theta_S\rangle$, (c) $\langle\cos^2\theta\rangle$ and (d)
  $\langle M^2\rangle$ for pyridazine in the presence of a static
  field $E_S=20$ kVcm$^{-1}$ and a laser field $I=10^{11}$ Wcm$^{-2}$ as
  a function of $\beta$ for the energetically lowest states with both $M+q+K$ an $K$
  even. The states are $0_{00}0$ (solid black), $1_{01}1$ (solid green),
  $2_{02}2$ (dash dotted   black), $1_{01}0$ (dash dotted green),
  $2_{02}1$ (dash black) and $2_{21}2$ (dash
  green).} 
\label{fig:pirazine_Static_20_beta_0_90_laser_1_10_11_mq_even_k_even}
\end{figure}

The inclination of the fields plays an essential role on their impact on the
rotational dynamics of the system, and as already discussed, the symmetries
are drastically modified as $\beta$ is varied.
Indeed, $\beta$ is another parameter in the Hamiltonian that enriches 
the physical phenomena observed, and its variation can provoke the appearance
of avoided crossings between energetically adjacent states of the same symmetry.
For pyridazine, we represent in Figs. 
\ref{fig:pirazine_Static_20_beta_0_90_laser_1_10_11_mq_even_k_even} (a) the
energy, (b) $\expected{\cos\theta_S}$, (c) $\expected{\cos^2\theta}$, and (d) $\expected{M^2}$, as a function
of $\beta$ with $E_S=20$ kVcm$^{-1}$ and $I=10^{11}$ Wcm$^{-2}$,
for the previous set of states with $K$ and $M+K+q$ even and 
following the same
labeling as in Sec. \ref{sec:increasing_laser}.
For these field strengths, the static electric field interaction dominates over
the laser field interaction, similar to 
Fig.~\ref{fig:pirazine_Static_0_200_laser_1_10_10_mq_even_k_even}.
In general, the energies show a smooth behavior as $\beta$ is varied, and 
depending on the state they exhibit an increasing or decreasing trend, see 
Fig.~\ref{fig:pirazine_Static_20_beta_0_90_laser_1_10_11_mq_even_k_even}(a),
e.g., the ground state energy increases from $-5$ cm$^{-1}$ to $-4.6$ cm$^{-1}$
from $\beta=0$ to $\pi/2$, respectively. 
The energy gap between the first four states is large enough to prevent the
presence of avoided crossings among them, the first avoided crossing being between 
the fourth and fifth excited states, $2_{02}1$ and $2_{21}2$, for
$\beta\approx 1.096$ (close to $3\pi/8$).
Regarding the orientation and alignment along the electric and laser fields
directions, respectively,  
different behaviors are observed. 
The ground state keeps a significant and approximately constant orientation
with $\expected{\cos\theta_S}>0.75$ for any value of $\beta$, as the electric
field is rotated away from the $Z$-axis the ground state probability density
follows this field. In contrast to this, since the laser interaction is not
dominant, its alignment is drastically reduced from
$\expected{\cos^2\theta}\approx 0.75$ to $0.22$ when $\beta$ increases from $0$ to
$\pi/2$.   
For any field configuration, the $1_{01}1$ level shows a moderate orientation
and alignment with a plateau-like behavior for $\expected{\cos\theta_S}$
and $\expected{\cos^2\theta}$. 
Compared to the ground state, the $2_{02}2$ level presents a similar
orientation and alignment for $\beta=0$, but a very different evolution
of these features as $\beta$ is varied, and for $\beta=\pi/2$, it keeps a
moderate alignment and a weak orientation. 
For parallel fields, $M$ is a good quantum number, and a non-parallel 
configuration allows the interaction and mixing between states with different
field-free $M$-value. This phenomenon is illustrated by means of the
expectation value $\expected{M^2}$ in 
Fig.~\ref{fig:pirazine_Static_20_beta_0_90_laser_1_10_11_mq_even_k_even}(d).
As the angle $\beta$ is increased, the evolution of $\expected{M^2}$ strongly
depends on the character of the corresponding level. 
For the ground state, $\expected{M^2}$ increases as $\beta$ is
enhanced, and for $\beta=\pi/2$, it reads $\expected{M^2}=0.746$.
The $\expected{M^2}$ value of the $1_{01}1$ level is close to 1 for $\beta\le
\pi/8$, but for larger values of $\beta$ decreases to
$0.6$ for $\beta=\pi/2$.
In contrast, for the other analyzed level with $M=1$,  
$2_{02}1$, the interaction with states with larger $M$ is dominant for
$\beta>\pi/4$, and $\expected{M^2}=3$ for $\beta=\pi/2$. 
For the considered $M=2$ levels, $2_{02}2$ and $2_{21}2$, the mixing with
states with lower $M$ is dominant, and $\expected{M^2}$ is smaller than in the
parallel configuration, e.g., for the $2_{02}2$ level and $\beta=\pi/2$, we have $\expected{M^2}=0.4063$. 

\begin{figure}
  \centering
  \includegraphics[scale=1.0,angle=0]{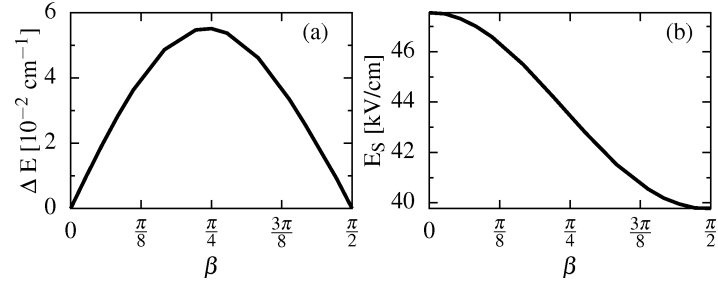}
  \caption{Width $\Delta E$ (a) and electrostatic field strength $E_S$ (b) at
the avoided crossing taking place between the states $2_{02}0$ and $2_{02}1$ for pyridazine, for $I= 10^{10}$ Wcm$^{-2}$ and different inclination
angles $\beta$.}
  \label{fig:avoided_crossing_following}
\end{figure}

With varying inclination angle $\beta$ between both fields the avoided crossings
leave their fingerprints in the relevant observable. For  pyridazine, the
states $2_{02}0$ and $2_{02}1$ belong to different irreducible representations
for $\beta=0$ and $\pi/2$, and to the same one for $0<\beta<\pi/2$. (Note that
the labeling of the states has been done in the same way as in
Fig.~\ref{fig:pirazine_Static_0_200_laser_1_10_10_mq_even_k_even}). 
For non collinear fields,
they suffer
an avoided crossing 
which  we have traced for $I=10^{10}$ Wcm$^{-2}$ and different values of
$\beta$ in  Fig. \ref{fig:avoided_crossing_following}. 
The results for the minimal energetical width  
$\Delta E=|E_{2_{02}0}-E_{2_{02}1}|$ and the electrostatic field strength 
at which this minimum appears are
presented in Figs. \ref{fig:avoided_crossing_following}(a) and (b),
respectively, as the angle $\beta$ is varied. 
For perpendicular fields, the levels suffer a real crossing and are
accidently degenerate $\Delta E=0$ for 
$I=10^{10}$ Wcm$^{-2}$ and $E_S=39.78$ kVcm$^{-1}$, 
whereas for $\beta=0$, they possess a different magnetic
quantum number $M$ and exhibit a symmetry-related crossing but
now for $E_S=47.55$ kVcm$^{-1}$.
As we see in panel (a), $\Delta E$ increases till it reaches the maximal value
of $5.33\cdot 10^{-2}$ cm$^{-1}$ for 
$\beta=\pi/4$, decreasing afterwards to $0$ for $\beta=\pi/2$.
The static field strength at which the avoided crossing takes places, see
panel (b), decreases monotonously as $\beta$ increases. 
Indeed, the variation of $E_s$ with $\beta$  is well-matched by the following
function $3.88\cos(2.06\beta)+43.67$,  and it is reduced by
 $7.74$ kVcm$^{-1}$ when $\beta$ goes from $0$ to $\pi/2$.

\section{Conclusions}
\label{sec:conclusions}

In this work, we have investigated the impact of a combination of an
electrostatic  and a nonresonant linearly polarized laser field
 on the rotational spectrum of asymmetric top molecules. 
This study has been performed in the framework of the Born-Oppenheimer
approximation considering that the vibrational and electronic dynamics
are not affected by the external fields. 
Our analysis is restricted to a rigid rotor description of molecules having
their permanent dipole moment parallel to one axis of inertia, and the
polarizability tensor is diagonal in the basis formed by the principal axis of
inertia.  
We have analyzed the symmetries and irreducible representations of the
Hamiltonian for the different field configurations. Numerically, each
irreducible representation has been treated independently, by expanding the
wave function in a basis respecting the corresponding symmetries. 
This procedure allows us to distinguish the avoided crossings from genuine
ones involving states of
the same and different symmetry, respectively. The presence of the avoided
crossings in the field-dressed spectrum affects the directional properties of
the molecule, they might significantly alter the
spectroscopy as well as the sterodynamic of the system. 
The labeling of a certain state and the passage through the avoided crossing that it suffers
depend on the way the symmetries are broken, i.e., on the temporal sequence
followed to turn on  the fields, which should be taken into account to
determine the adiabaticity of a certain process. 

The richness and variety of the resulting field-dressed rotational dynamics
have been illustrated by analyzing the energetic shifts, as well as the
orientation, alignment and the hybridization of the angular motion.
As prototype example we have investigated the fluorobenzene and pyridazine
molecules.   
For three field configurations, the evolution of a certain set of
states belonging to a certain symmetry has been analyzed with varying
electrostatic field strength or laser intensity. Different types of
behavior were observed, depending on the dominant field interaction as well
as on the considered molecular system, through its rotational constant, dipole
moment and polarizability tensor.  
Due to the competition between both interactions, the features of the rotational
spectrum are significantly changed as the field parameters are modified.  
In the strong laser field regime, the presence of an intense electric field
reduces the orientation of the ground state, especially as $\beta$ is rotated
from zero to larger values,
and highly excited levels only present a very weak alignment.
Whereas, if the electrostatic field is dominant the states are oriented along
its direction, and they only present a mild alignment along the $Z$-axis in
the LFF.
We have shown that a proper combination of non-collinear fields gives rise to 
a strong orientation along the static field direction together with a
2D-alignment on the other two axis of the molecule.
In particular for $\beta=\pi/2$, the molecular plane is fixed onto the
$XY$-plane of the LFF, and the orientation is along the $X$ axis. 
Finally, we have also investigated the role played by the inclination angle of
the fields $\beta$, by analyzing the spectral properties of several states, the
loss of the azimuthal symmetry has been quantified by the expectation value
$\langle M^2\rangle$, which is a conserved magnitude for parallel fields. 
An avoided crossing between two states has been traced as $\beta$ is
modified, the electrostatic field strength at which it takes places varies
within a few kVcm$^{-1}$, and the corresponding energetical width will
allow us to compute the adiabaticity of the crossing once the variation of
the field strength is known. A natural extension to this work would be to
consider other molecular systems, especially different conformers of the same
molecule, looking for specific phenomena that might help to distinguish
between the molecules. 

\begin{acknowledgments}
Financial support by the Spanish project FIS2008-02380 (MICINN) as well as the
Grants FQM-2445 and FQM-4643 (Junta de Andaluc\'{\i}a) is gratefully
appreciated. J.J.O. and R.G.F. belong to the Andalusian research group FQM-207. 
J.J.O. acknowledges the support of ME under the program FPU.
We thank J. K\"upper for fruitful discussions. 

\end{acknowledgments}

\appendix
\section{Wigner and Hamiltonian matrix elements}
\label{sec:matrix}
The field-free eigenstates of a symmetric top molecule, see eq.
\eqref{eqn:sym_basis}, are proportional to the Wigner
matrix elements, $D_{M,K}^J(\Omega)$, which are defined as 
\begin{equation}
 D_{M,K}^J(\Omega)=e^{-iM\phi}d_{M,K}^J(\theta)e^{-iK\chi},
\label{eqn:wigner_matrix}
\end{equation}
where $d_{M,K}^J(\theta)$ are the reduced Wigner matrix elements~\cite{zare}.
% that can be calculated as
%\begin{eqnarray*} 
%d^J_{M,K}(\theta)&=&\sqrt{(J+M)!(J-M)(J+K)(J-K)!}\\
%&&\times \sum_\nu\cfrac{(-1)^{M-K+\nu}\sin\left(\frac{\theta}{2}\right)^{M-K+2\nu}\cos\left(\frac{\theta}{2}\right)^{2J+M-K+2\nu}}{(J+K-\nu)!\nu!(K-M+\nu)!(J-K-\nu)!},
%\end{eqnarray*}
%where $\nu$ is integer and satisfies that the arguments of the factorial are
%greater than zero. 
To evaluate the matrix elements of the Hamiltonian, we have used the following
properties of the Wigner matrix,  the  complex conjugate
    \begin{equation}
      D^{J\dagger}_{M,K}(\Omega)=(-1)^{M-K}D^J_{-M,-K}(\Omega),
    \end{equation}
the relation between the Wigner reduced matrix elements reads
    \begin{eqnarray}
      d^j_{m',m}(\theta)&=& (-1)^{m-m'}d^j_{m,m'}(\theta)\\
      d^j_{m',m}(\theta)&=& (-1)^{m'-m}d^j_{-m',-m}(\theta)\\
      d^j_{m',m}(\theta)&=& d^j_{m,m'}(-\theta)
    \end{eqnarray}
and the integral of the triple product of Wigner matrices
\begin{eqnarray}
\nonumber
&&\int\;d\Omega D_{-K,-M}^J(\Omega)D_{K_1,M_1}^{J_1}(\Omega)D_{K_2,M_2}^{J_2}(\Omega)=\\
&=& 8\pi^2
\begin{pmatrix}
J_1 & J_2 & J \\
M_1 & M_2 & -M\\
\end{pmatrix}
\begin{pmatrix}
J_1 & J_2 & J \\
K_1 & K_2 & -K\\
\end{pmatrix}.
\label{eqn:int3}
\end{eqnarray}
where $
\begin{pmatrix}
j_1 & j_2 & j_3 \\
m_1 & m_2 & m_3\\ 
\end{pmatrix}
$ are the \emph{3J Symbols}.

For completeness, we provide the non-zero matrix elements appearing in the
evaluation of the full  Hamiltonian. 
For the  field-free Hamiltonian  $H_r$ \eqref{eqn:hamiltonian_rot},
we have  
\begin{align}
\nonumber
 \bra{J'K'M'}H_r\ket{JKM}&=(AJ(J+1)+CK^2)\delta_{J',J}\delta_{K',K}\delta_{M',M}\\
\nonumber
&+B\sqrt{J(J+1)-K(K+1)}\sqrt{J(J+1)-(K+1)(K+2)}\delta_{J',J}\delta_{K+2,K}\delta_{M',M},
\end{align}
where $A$, $B$ and $C$ are defined as $A=(B_x+B_y)/2,\quad
B=(B_x-B_y)/4,\quad\text{and}\quad C=(2B_z-B_x-B_y)/2$~\cite{kroto}.
The Stark interaction $H_S$ \eqref{eqn:hamiltonian_static}
rewritten in terms of Wigner
matrix elements, is given by
\begin{equation}
\label{eqn:h_static_wigner2}
 H_S=-\mu E_S\cos\theta_S=-\mu E_S h_S= -\mu E_S\left(\cos\beta
 D_{00}^1(\Omega)+\sin\beta\sqrt{\frac{1}{2}}(D_{-1,0}^1(\Omega)-D_{10}^1(\Omega))\right), 
\end{equation}
Thus, the non-zero matrix elements are
\begin{eqnarray}
\nonumber
\bra{JKM}h_S\ket{JKM}&=&\frac{MK}{J(J+1)}\cos{\beta}\\
\nonumber
&&\\
\nonumber
\bra{JKM+1}h_S\ket{JKM}&=&\frac{K\sin{\beta}}{2J(J+1)}\sqrt{J(J+1)-M(M+1)}\\
\nonumber
&&\\
\nonumber
\bra{J+1KM}h_S\ket{JKM}&=&\cos\beta\sqrt{\frac{[(J+1)^2-M^2][(J+1)^2-K^2]}{(2J+3)(2J+1)(J+1)^2}}\\
\nonumber
&&\\
\nonumber
\bra{J+1KM+1}h_S\ket{JKM}&=& -\frac{\sin{\beta}\sqrt{[(J+1)^2-K^2](J+M+1)(J+M+2)}}{2(J+1)\sqrt{(2J+3)(2J+1)}}\\
\nonumber
&&\\
\nonumber
\bra{J+1KM-1}h_S\ket{JKM}&=&-\bra{J+1K-M+1}h_S\ket{JK-M}
\end{eqnarray}
The laser Hamiltonian $H_L$  \eqref{eqn:hamiltonian_static} takes now the form
\begin{equation}
H_L=-\cfrac{ I}{2\epsilon_0c}h_{L}
%=-\cfrac{\langle E_L^2\rangle}{2}
%\left(\alpha^{zx}\cos^2\theta+\alpha^{zy}\sin^2\theta\sin^2\chi\right)
= -\cfrac{I}{2\epsilon_0c}\left(\frac{\alpha^{zx}+\alpha^{yx}}{3}D_{00}^2(\Omega)-\frac{\alpha^{yx}}{\sqrt{6}}[D_{02}^2(\Omega)+D_{0-2}^2(\Omega)]+\cfrac{\alpha^{zx}+\alpha^{yx}}{3}\right).
\label{eqn:h_laser_wigner2}
\end{equation}
In this expression  we have omitted the terms which only
introduce a shift in the energy. Performing the integrals corresponding to
each term, we get
\begin{eqnarray}
\nonumber
\bra{JKM}h_L\ket{JKM}&=&\left(\frac{\alpha^{zx}+\alpha^{zy}}{3}\right)\cfrac{[3M^2-J(J+1)][3K^2-J(J+1)]}{2J(J+1)(2J-1)(2J+3)}+\cfrac{\alpha^{zx}+\alpha^{yx}}{3}\\
\nonumber
&&\\
\nonumber
\bra{J+1KM}h_L\ket{JKM}&=& (\alpha^{zx}+\alpha^{zy})\frac{MK\sqrt{[(J+1)^2-M^2][(J+1)^2-K^2]}}{J(J+1)(J+2)\sqrt{(2J+3)(2J+1)}}\\
\nonumber
\bra{J+2KM}h_L\ket{JKM}&=&\left(\frac{\alpha^{zx}+\alpha^{zy}}{2}\right)\sqrt{[(J+2)^2-K^2][(J+1)^2-K^2]}\\
\nonumber
&&\times\frac{\sqrt{[(J+2)^2-M^2][(J+1)^2-M^2]}}{(J+1)(J+2)(2J+3)\sqrt{(2J+1)(2J+5)}}\\
\nonumber
\end{eqnarray}
\begin{eqnarray}
\nonumber
\bra{JKM}h_L\ket{JK+2M}&=&-\alpha^{yx}[3M^2-J(J+1)]\\
\nonumber
&&\times\frac{\sqrt{[J^2-(K+1)^2](J-K)(J+K+2)}}{2J(J+1)(2J+3)(2J-1)}\\
\nonumber
&&\\
\nonumber
\bra{J+1KM}h_L\ket{JK+2M}&=&-\alpha^{yx}M\sqrt{[(J+1)^2-M^2]}\\
\nonumber
&&\times\frac{\sqrt{[(J-K)^2-1](J-K)(J+K+2)}}{2J(J+1)(J+2)\sqrt{(2J+1)(2J+3)}}
\end{eqnarray}
\begin{eqnarray}
\nonumber
\bra{J+2KM}h_L\ket{JK+2M}&=&-\alpha^{yx}\sqrt{[(J+2)^2-M^2][(J+1)^2-M^2]}\\
\nonumber
&&\times\frac{\sqrt{[(J-K)^2-1](J-K)(J-K+2)}}{4(J+1)(J+2)(2J+3)\sqrt{(2J+1)(2J+5)}}\\
\nonumber
&&\\
\nonumber
\bra{J'K'M'}h_L\ket{JKM}&=&\bra{J'-K'-M'}h_L\ket{J-K-M}.
\nonumber
\end{eqnarray}
%\bibliographystyle{apsrev}
%\bibliographystyle{aipsamp}
%\bibliography{asymmetric_biblio}
%

\end{document}